\documentstyle[epsf]{elsart}

\begin{document}
\begin{frontmatter}

\title{Evolution of the System \\
with Singular Multiplicative Noise}
\author{Alexander I.~Olemskoi
and Dmitrii O.~Kharchenko
\thanksref{AODIKH}
}
\address{Sumy State University\\ 2, Rimskii-Korsakov
St., 244007 Sumy UKRAINE
}
\thanks[AODIKH]{alexander\char'100olem.sumy.ua\\ ~~~~~dikh\char'100ssu.sumy.ua}

\begin{abstract}
The governed equations for the order parameter,
one-time  and two-time correlators
are obtained on the basis of the
Langevin equation with the white multiplicative noise
which amplitude $x^{a}$ is determined by an exponent $0<a<1$
($x$ being a stochastic variable).
It  turns out that equation for autocorrelator
includes an anomalous average of the
power-law function with the fractional exponent $2a$.
Determination of this average for the stochastic system with a
self-similar phase space is performed. It is shown that  at $a>1/2$,
when the system is disordered, the correlator behaves
non-monotonically in the course of time, whereas the autocorrelator is
increased monotonically.  At $a<1/2$ the phase portrait of the system
evolution divides into two domains: at small initial values of the
order parameter, the system evolves to a disordered state, as above;
within the ordered domain it is attracted to the point having the
finite values of the autocorrelator and order parameter. The
long-time asymptotes are defined to show that, within the disordered
domain, the autocorrelator decays hyperbolically and the order
parameter behaves as the power-law function with fractional exponent
$-2(1-a)$.  Correspondingly, within the ordered domain, the behavior
of both  dependencies is exponential with an index proportional to
$-t\ln t$.

\vspace{0.5cm}


\vspace{0.5cm}


\end{abstract}

\end{frontmatter}
\section{Introduction}

Within the framework of the ordinary thermodynamic approach, it is
postulated that the bath is passive with respect to a variation of
the system state parameter $x$ which is an amplitude of the
hydrodynamic mode [1]. In such a case the noise of the corresponding
stochastic process $x(t)$ is additive one in the course of time $t$
to have the temperature as intensity, which is independent
on the variable $x$. On the contrary, the amplitude $g(x)$ of a
multiplicative noise varies with the stochastic variable $x$. Such
type examples represent a population dynamics [2], directed
percolation [3], L\'evy flights [4] and so on.

According to Ref.[5], the multiplicative function $g(x,t)$ represents
the homogeneous function for the systems with a self-similar phase
space. In such a case we can write
the noise amplitude as the power-law
function  \begin{equation} g(x)=x^a, \qquad
a\in[0,~1]\label{8} \end{equation} that is singular in character
because $g=0$ at $x=0$.  This kind assumption allows us to consider
such models as ordinary thermodynamic system with an additive noise
$(a=0)$, directed percolation process $(a=1/2)$, population dynamics
$(a=1)$ and so on.

This work is organized as follows. In Sec.2 the governed equations
for the first moment (order parameter) of the stochastic variable
$x(t)$ as well as the one-time and two-time correlators are
obtained on the basis of the Langevin equation with the white
multiplicative noise. It turns out that equation for
autocorrelator gains an anomalous average of the squared power
function (\ref{8}) with fractional exponent $2a$ (the fractional
average). Sec.3 deals with the determination of this average for the
stochastic system with a self-similar phase plane, whose
distribution function is a homogeneous one and can be approximated by
a power function [5]. Within the framework of this approach, the
system behavior is governed by the value of the exponent $a$ in
Eq.(\ref{8}) [5,6].
At $a>1/2$ the system is disordered to be represented by correlator
and autocorrelator (Sec.4). At $a<1/2$ the evolution of the order
parameter ought to take into account (Sec.5). Sec.6 contains a short
conclusions.

\section{Basic equations}

As usually, let us start with the Langevin equation
\begin{equation}
{{\rm d}x\over{\rm d}t }=f(x)+g(x)\xi(t)\label{1}
\end{equation}
for a stochastic variable $x(t)$.
In right-hand side of Eq.(\ref{1}), $f(x)$ is  the deterministic
evolution force, the second term defines the multiplicative noise
with the amplitude $g(x)$.  The statistical properties of the
Langevin force $\xi(t)$ are standard:  \begin{equation}
\langle\xi(t)\rangle=0,\qquad
\langle \xi(t)\xi(t')\rangle=\delta(t-t')
\label{2}
\end{equation}
where the angle brackets denote the averaging.
In order to study the simplest system properties,
we consider the order parameter
$\eta(t)\equiv\langle x(t)\rangle$,
the two-time correlator
$G(t,t')\equiv\langle \delta x(t) \delta x(t')\rangle$,
$\delta x(t)\equiv x(t) - \langle x(t)\rangle$
and the autocorrelator (structure factor)
$S(t)\equiv\langle (\delta x(t))^2\rangle$.
Within the framework of the white-noise approximation
that is expressed by  Eqs.(\ref{2})
and the Ito calculus, we can treat $g(x)$ and $\xi(t)$ as statistically
independent functions.
Then, at averaging of Eq.(\ref{1}) we can set $\langle
g(x)\xi(t)\rangle=0$  so that the evolution of the first moment
is defined by the equation
$\dot\eta = \langle f(x)\rangle$
where $\langle f(x)\rangle\ne f(\eta)$, dot stands for the derivative with
respect to the time $t$.
Without loss of
generality, the deterministic part of the evolution force can be
chosen in the polynomial Landau form:  \begin{equation}
f(x)=-{\partial V(x)\over \partial x},\qquad V(x)=-{\varepsilon
\over 2}~x^2 +{1\over 4}~x^4 \label{3} \end{equation} where
$-\varepsilon$ is an external driven parameter type of the
dimensionless temperature counted from a critical value.
Performing the averaging under the force
definition (\ref{3}), we get term $\langle x^3\rangle$ that
is reduced to $\eta\langle x^2\rangle\equiv \eta (\eta^2 + S)$, in
accordance with the cumulant expansion.  As a result, the evolution
equation for order parameter takes the form  \begin{equation} \dot
\eta=\eta\left(\varepsilon-\eta^2\right)-3\eta S\label{4}
\end{equation} where the values $\eta$, $S$ are dependent on the time
$t$.

By  analogy, accounting the equation
$$\langle g(x(t))x(t')\xi(t)\rangle=
\langle g(x(t))x(t')\rangle \langle \xi(t)\rangle+
\langle g(x(t))\xi(t)\rangle \eta(t')+
\langle g(x(t))\rangle \langle x(t')\xi(t)\rangle=0$$
that follows from Eq.(\ref{2}) and above-mentioned property of the white noise,
we obtain the equation for two-time correlator
\begin{equation}
{\partial \over \partial t}
G(t,t')=\lbrace\varepsilon-3[\eta^2(t)+S(t)]\rbrace G(t,t').\label{5}
\end{equation}

The problem lies now in obtaining an evolution equation
for the variance $S \equiv\langle x^2\rangle - \eta^2$.
It can be performed
if we use the relation ${\rm d}x^2\equiv (x+{\rm d}x)^2-x^2$
where, in accordance with Eq.(\ref{1}),
the differential ${\rm d}x$ is written as follows:
\begin{equation}
{\rm d}x=f(x){\rm d}t+g(x){\rm d}w, \quad {\rm d}w\equiv\xi(t){\rm
d}t, \quad ({\rm d}w)^2={\rm d}t.  \label{6} \end{equation}
Then, up to the first derivative, the equation
for $\langle x^2\rangle$ takes
the form \begin{equation} {{\rm d} \over {\rm d}t}
\langle x^2\rangle= 2\langle xf(x)\rangle+\langle
 g^2(x)\rangle.\label{7} \end{equation} Here the last term is the
average intensity of the multiplicative noise being a result of the
interaction of the variable $x$ with a bath, whose variables have
been appropriately eliminated.
Inserting definition given by Eq.(\ref{8}) into
Eq.(\ref{7}), we come up against the average $\langle
x^{2a}\rangle$ of the stochastic variable with a fraction exponent $2a$.
Further, we are coming to the obtaining an
expression for such fractional average in
terms of the cumulant expansion.

\section{Calculation of the fractional average}

Let us admit that, apart from the initial distribution
$P(x)$, there is another distribution $P_q(x)$
specified by a positive parameter $q<1$.
Moreover, for $P(x)$ and $P_q(y)$ we assume the following relation:
\begin{equation}
x^qP(x)dx\equiv yP_q(y)dy.\label{9}
\end{equation}
Then, the probability density $P_q(y)$
for the new stochastic variable $y\equiv x^q$
can be rewritten as the normalized distribution
\begin{equation}
P_q(y)=q^{-1}y^{(1-q)/q}P(y^{1/q}).
\label{10}
\end{equation}
Denoting the averaging over the distribution $P_q(y)$ as
$\langle \ldots \rangle_q$ and using the designation
$\langle \ldots \rangle$ for the initial distribution $P(x)$,
we obtain
\begin{eqnarray}
&&\langle x^q\rangle = \langle y\rangle_q; \label{11}
\\ \langle x^q\rangle\equiv\int x^qP(x){\rm d}x,&&\qquad
\langle y\rangle_q\equiv
q^{-1}\int y y^{(1-q)/q}P(y^{1/q}){\rm d}y.\nonumber
\end{eqnarray}
Thus, the distribution (\ref{10})
allows us to use the usual cumulant expansion
 for any average $\langle x^q\rangle$ with the
fractional exponent  $q$.

For self-similar stochastic systems
the distribution function can be written
in  a power-law form
\begin{equation}
P(x)\simeq A x^{-2a},\qquad A\equiv {1\over
2}|1-2a|~b^{|1-2a|}\label{12} \end{equation} to be a homogeneous
function [5], [7]. Here factor $A$ is
responsible for a  cut-off procedure in normalization condition
\begin{equation}
2\int\limits_b^{1/b}P(x){\rm d}x=1
\end{equation}
with cut-off parameter $b\to 0$.  The integrating with the distribution
function (\ref{12}) gives
\begin{eqnarray}
\langle x^{nq}\rangle\equiv A\int
x^{nq}x^{-2a}{\rm d}x=A(1-2a+nq)^{-1} x^{1-2a+nq}, \\
\langle x^n\rangle \equiv A \int x^n x^{-2a}{\rm d}x
=A(1-2a+n)^{-1}  x^{1-2a+n}.
\end{eqnarray}
As a result, using Eq.(\ref{11}), we obtain the relation
\begin{equation} \langle
x^{nq}\rangle =\alpha_n(q)\langle x^n\rangle^{p_n(q)}\label{13}
\end{equation} where the exponent $p_n(q)$ and the multiplier
$\alpha_n(q)$ are introduced as follows:  \begin{equation}
p_n(q)={1-2a+nq \over 1-2a+n},\qquad \alpha_n(q)=A^{n(1-q)\over
(1-2a+n)}~p_n^{-1}(q)~(1-2a+n)^{p_n(q)-1} .\label{14} \end{equation}

Now, we are ready to formulate the equation for the autocorrelator
$S=\langle x^2\rangle-\eta^2$ on the basis of Eqs.(\ref{7}),
(\ref{8}), (\ref{13}), (\ref{14}). According to  [6] a keypoint of
the system with the multiplicative noise is that its behavior is
governed by the magnitude of the exponent $a$ in Eq.(\ref{8}).
At $1/2<a<1$, when the fractal
dimension of the phase space $D=2(1-a)$ is less than 1, the system is
always disordered and its evolution is represented by
the correlator $G(t,t')$ and the structure factor $S(t)$. The former
is governed by Eq.(\ref{5}), for the latter it follows
\begin{eqnarray}
&\dot S=2S(\varepsilon-3S)+\alpha_2S^{p_2},& \label{15}\\
&\alpha_2\equiv\alpha_2(a)=A^{2(1-a)p_2}p_2^{-p_2},\qquad p_2\equiv
p_2(a)=(3-2a)^{-1}&\nonumber \end{eqnarray}
from Eqs.(\ref{4}), (\ref{7}), (\ref{8}), (\ref{13}), (\ref{14}) if
we put $q=a$, $n=2$.
Within another domain $0<a<1/2$, the above fractal dimension $D>1$
so that the system can be ordered
and instead of Eq.(\ref{15}) we obtain
\begin{eqnarray}
&\dot S =2 S \left[\varepsilon-3(\eta^2+S)
\right]+\alpha_1\eta^{p_1},&\label{16}\\
&\alpha_1\equiv\alpha_1(2a)=A^{(1-2a)p_1}p_1^{-p_1},\qquad p_1\equiv
p_1(a)=[2(1-a)]^{-1}&\nonumber
\end{eqnarray}
at $q=2a$, $n=1$ in Eqs.(\ref{13}), (\ref{14}).

\section{Evolution of disordered system}

As pointed out above, in the case  when the exponent $a>1/2$
(the fractal dimension $D<1$),
the system is governed by Eqs.(\ref{15}), (\ref{5}) for the one-time
and two-time correlators $S(t)$, $G(t,t')$ being the structure factor
and Green response function. The form of the time-dependence for the
former is shown in Fig.1a.  It is  seen that $S(t)$ monotonically
increases to the stationary magnitude $S_0$
 determined by the equation
\begin{equation}
\varepsilon-3 S_0+(\alpha_2/2)S_0^{p_2-1}=0.\label{17}
\end{equation}
In the limit $S\ll 1$ when $S^{p_2}\gg S\gg S^2$,
Eq.(\ref{15}) gives the power-law time dependence
\begin{equation} S(t)=
\left({A^{2(1-a)}\over p_2 (1-p_2)}\right)^{p_2/(1-p_2)}t^{1/(1-p_2)},\qquad
p_2\equiv(3-2a)^{-1}
\label{18} \end{equation}
where we put $S(t=0)=0.$ In opposite case $S_0-S\ll S_0$ one has the
exponential dependence $S-S_0\propto e^{-\lambda t}$, $\lambda\equiv
6(2-p_2)S_0-2(1-p_2)\varepsilon$. According to Eq.(\ref{17}) the
stationary value $S_0$ raises with $\varepsilon$ increase from the
minimal magnitude $(\alpha_2/6)^{1/(2-p_2)}$ (see Fig.1b).

The solutions of Eq.(\ref{5}) for different values  $a$ and
$\varepsilon$ are shown in Fig.2.  We plot correspondent
dependencies of the Green function $G(t,0)$ at identical initial
conditions. It is interesting to observe that the correlator
$G(t,0)$ reaches firstly its maximum and then monotonically decreases
to zero.  The function $G(t,0)$
attains its maximum  more sharply if
the parameter $\varepsilon$ increases (cf. curves 1, 2).  The
exponent $a$ increasing drives to the similar effect (cf. curves 1,
3).

\section{Evolution of ordering system}

Now one has $a<1/2$, $D>1$ and the system behavior is governed by
Eqs.(\ref{4}), (\ref{5}), (\ref{16}), from which the first and
third state the enclosed system of differential equation. To analyze
the latter, it is convenient to use the phase plane method. As is seen
from the corresponding phase portrait in Fig.3a, at small values
$\varepsilon$ there is only one attractive point $\eta_0=0$,
$S_0=\varepsilon/3$ (Fig.3a). With $\varepsilon$ increase at the point
\begin{equation} \varepsilon_0={4-p_1\over 2-p_1}
\left[ {3\over 8} (2-p_1)\alpha_1\right]^{2/(4-p_1)}\label{19}
\end{equation}
a bifurcation creates new saddle and attractive  points (see Fig.4)
with coordinates $\eta_c=[(2-p_1)(4-p_1)^{-1}\varepsilon_0]^{1/2}$,
$S_c=(2/3)(4-p_1)^{-1}\varepsilon_0$.  It is seen from Fig.5 that
bifurcation temperature $\varepsilon_0$ is increased infinitely with
the  exponent  $a$ growth to the critical value  $a=1/2$. The
coordinates of the new stationary points are determined by equations
\begin{equation}
\varepsilon-\eta_0^2-(3/4)\alpha_1\eta_0^{p_1-2}=0,\qquad
S_0=(4\alpha_1)^{-1}(\varepsilon-3S_0)^{(p_1/2)-1},\label{20}
\end{equation}
that is obtained from Eqs.(\ref{4}), (\ref{16}) at $\dot \eta=0$,
$\dot S=0$. The corresponding dependencies $\eta_0(\varepsilon)$,
$S_0(\varepsilon)$ are depicted in Fig.4 where the dashed curves are
respective for the saddle $S$ and solid ones --
for the attractive point $C$.

It is interesting to note that the system undergoes the phase
transition of the first order, despite of the bare $x^4$-potential
(\ref{3}) corresponds to the continuous one. Thus, at small values
of the exponent of the multiplicative noise (\ref{8}) $(a<1/2)$ the
fluctuations transform order of the phase transition, whereas at
$a>1/2$ ones suppress the ordering process at all.

Let us return now to analysis of the time dependencies
of the main averages under consideration.
Firstly, we analyze the system evolution to the disordered state
(a vicinity of the point $C_0$ in Fig.3)
for extremely large time $t\to\infty$.
The keypoint of our consideration is that
the fractional average appearance in Eq.(\ref{16})
does not allow to use the ordinary Lyapunov`s method
because the exponential time-dependence becomes invalid.
Instead, let us introduce the generalized exponential form:
\begin{equation}
e^{qt}\to E_{ q}(t)\equiv [1 + (1 - q) t]^{1/1-q}
\label{a0}
\end{equation}
where $ q$ is a generalized Lyapunov index.
First, such a type of generalization was used by Tsallis [8]
to obtain the ordinary Gibbs-Boltzmann exponent in the limit
$ q\to 1$.
In our case, the latter is arbitrary so that
function $E_{ q}(t)$ acquires the power-law character,
in particular the follow derivation rule is fulfilled:
\begin{equation}
\partial E_{ q}(t)/\partial t =\left(E_{ q}(t)\right)^q
\equiv E_{q}^q (t).
\label{b0}
\end{equation}
In the limits of the short and long times this function has the
asymptotic behavior:  \begin{equation} \lim_{t\to 0}E_{ q}(t)\to
1+t,\qquad \lim_{t\to\infty}E_{ q}(t)\to \left((1-q)t\right)^{1/1-q}.
\label{c0}
\end{equation}
Below, the first of the asymptotes will be used for extraction of the
Lyapunov-type multipliers, the second one allows us to set an index
$q$ (see after Eq.(\ref{a4})).

Let us define the solutions of Eqs.(\ref{4}), (\ref{16})
in the form
\begin{equation}
\eta(t)=mE_{\mu}(t), \quad S(t)=S_0+nE_{\nu}(t)
\label{a1}
\end{equation}
where $S_0=\varepsilon/3$ to correspond to the point $C_0$;
the exponents $\mu$, $\nu$ and coefficients $m, n$
must be determined.
\footnote{It is worth to note that the parameter $n\ll 1$ to be a 
second term
in expansion (\ref{a1}), whereas a magnitude $m$ can be arbitrary because it 
stands as a first term there. Physically, it means that the order parameter
behaves in non-linear manner, in contrast to a linear regime of
the  structure factor.} 
Inserting  Eqs.(\ref{a1}) into Eqs.(\ref{4}),
we obtain up to the first power of $m, n\ll 1$
\begin{equation}
3nE_{\mu}^{1-\mu}(t)E_{\nu}(t)=-1.
\label{a2}
\end{equation}
From this, within the long-time approximation (\ref{c0}), one obtains
\begin{equation}
n^{-1}=3(1-\mu),\quad\nu=2.
\label{a3}
\end{equation}
Respectively, the inserting Eq.(\ref{a1}) into
Eq.(\ref{16}) gives
\begin{equation}
E^{1-\nu}_{\nu}(t)
\left [2\varepsilon n
-\alpha_1 m^{p_1}E^{p_1}_{\mu}(t)E^{-1}_{\nu}(t)\right ]=-n.
\label{a4}
\end{equation}
As has been pointed out above,
in the short-time limit
the function $E^{1-\nu}_{\nu}(t)$ can be taken as
$1$ to correspond to extraction of the Lyapunov
multiplier.
Respectively, in
the long-time limit there is
$E^{p_1}_{\mu}(t)E^{-1}_{\nu}(t)={\rm const}\equiv p_1^{-1}$ and we
obtain
\begin{equation} 3\alpha_1
m^{p_1}=-(1+2\varepsilon), \quad\mu=1+p_1\equiv 1+[2(1-a)]^{-1}.
\label{a5} \end{equation} Thus, within  the long time-approximation,
the structure factor \begin{equation}
S(t)=S_0+(2/3)(1-a)t^{-1},\qquad t\to\infty\label{_29}
\end{equation}
tends to
the stable magnitude $S_0$ hyperbolically. The order
parameter is decreased according to the power dependence
\begin{equation}
\eta(t)=\eta_0-[2(1-a)]^{2(1-a)}|m|t^{-2(1-a)},\qquad
t\to\infty\label{_30} \end{equation}
with the
exponent decreasing with the parameter $a$; the amplitude $m$
is given by the first equation (\ref{a5}).

Thus, in accordance with the phase portraits in Fig.3a, at
$\varepsilon<\varepsilon_0$ the order parameter $\eta(t)$ decreases
monotonically in the course of time, whereas the structure factor can
vary non-monotonically, in contrast to the case $a>1/2$ (cf. Fig.1).
More complex behavior appears within the domain
$\varepsilon>\varepsilon_0$ when  the
ordered state occurs due to the bifurcation.
As is seen from Fig.3b the phase plane
divides into two domains corresponding to small and large values of
the order parameter.  Within the former, the system behaves as at
the above case $\varepsilon<\varepsilon_0$, but if an initial magnitude
of the order parameter is more a critical value,
the system passes to the attractive point $C$. The corresponding
dependencies of the time are depicted in Fig.6. It is characterically
that these dependencies display a  critical slowing-down near the
separatrix $C_0SC$ in Fig.3b (see curves 1, 2 in Figs.6a, 6b).

In order to analyze the long-time behavior in the vicinity of the
point $C$, we can not use the solution like the generalized exponent
(\ref{a0}). The latter is applicable when nonlinearity effects are
sufficient to fix the above mentioned amplitudes $m$, $n$ by
Eqs.(\ref{a3}), (\ref{a5}). In the case under consideration, the
linear conditions are satisfied and instead of the generalized
exponent $(\ref{a0})$ we ought to use the Mellin transformation. The
principle difference  is that the former is defined by the single
index $ q$, whereas the later contains a set of $ q$.
Inserting the definitions (cf. Eqs.(\ref{a1}))
\begin{equation}
\eta(t)=\eta_0+\int m_q t^q{\rm d}q,
\label{q1}
\end{equation}
\begin{equation}
S(t)=S_0+\int n_q t^q {\rm d}q
\label{q2}
\end{equation}
into Eqs.(\ref{4}), (\ref{16}) being written
within the linear approximation, we obtain
the equations for  specific amplitudes $m_q, n_q\ll 1$:
\begin{equation}
(q/t+2\eta_0^2)m_q-3\eta_0n_q=0,
\label{q3}
\end{equation}
\begin{equation}
[4\eta_0(\varepsilon-\eta_0^2)-\alpha_1\eta_0^{p_1-1}p_1]m_q+
[q/t+2(\varepsilon+\eta^2)]n_q=0.
\label{q4}
\end{equation}
This system has solutions provided the ratio $c\equiv -q/t$
is determined by equation
\begin{equation}
c=(\varepsilon+\eta^2_0)\left[1\pm\sqrt{1-
{8\eta_0^2(2\varepsilon-\eta_0^2)-3\alpha_1\eta^{p_1}p_1\over
\varepsilon+2\eta^2_0}}\right].
\label{q5}
\end{equation}
Thus, in a vicinity of the ordered point $C$ the dependencies of the
order parameter and structure factor (\ref{q1}), (\ref{q2}) take the
forms
\begin{equation}
\eta(t)=\eta_0+m\exp\left(-c t\ln t\right),
\label{q6} \end{equation}
\begin{equation}
S(t)=S_0+n\exp\left(-c t\ln t\right),
\label{q7}
\end{equation}
where the amplitudes $m$, $n$ correspond to the index $ q=-ct$.

Finally, the time dependencies of the Green function
$G(t,0)$ are determined by Eqs.(\ref{4}), (\ref{5}), (\ref{16}) as is
shown in Fig.7. At $\varepsilon<\varepsilon_0$, the monotonic decrease
appears for large value of initial magnitude $\eta(0)$ of the
order parameter (see Fig.7a).
In the case $\varepsilon>\varepsilon_0$, a maximum of
dependence $G(t,0)$ disappears at condition $\eta(0)>\eta_c$ corresponding
to the domain of the ordering state (see Fig.7b).

\section{Conclusion}

Summarizing, we have derived and analized the equations for order parameter,
structural factor and correlation function to describe the evolution
of the system with multiplicative noise. In representation of the
noise amplitude as the power-low function $g(x)=x^a$, we have
discussed a way to apply the cumulant expansion for average
$\langle x^{2a}\rangle$ in the case of a self-similar phase space.

We have shown that the system behavior is given by magnitude of the
exponent $a$: at $a>1/2$  the system  is disordered; at $a<1/2$
a phase transition to the ordered state is observed. In the first case
the obtained time dependencies for the structure factor show the
monotonic increasing to a stationary state, whereas the correlator
shows the non-monotonic behavior. In the second case $(a>1/2)$
we would have to investigate the system behavior on the phase  plane
given by order parameter $\eta$ and structure factor $S$.
It was shown that the phase portrait falls into two domains
characterized by disordered and ordered states
(the former corresponds to small
initial values of $\eta$, the latter -- to finite
values of both $\eta$ and $S$).  We have
found how the system attains the steady states at long-time asymptotics.
Within disordered domain, the structure factor decays
hyperbolically and the order parameter dependence is described by the
power-law function with the exponent $-2(1-a)$.
In the ordered phase, the order parameter and structure factor
exhibit
exponential behavior with index proportional to $-t\ln t$.

\section{Acknowledgments}

Authors are grateful to Charles University and Institute of Physics
(Prague, Czech Republic) for a hospitality that allows to perform
this work.

\newpage
\centerline{Captions}

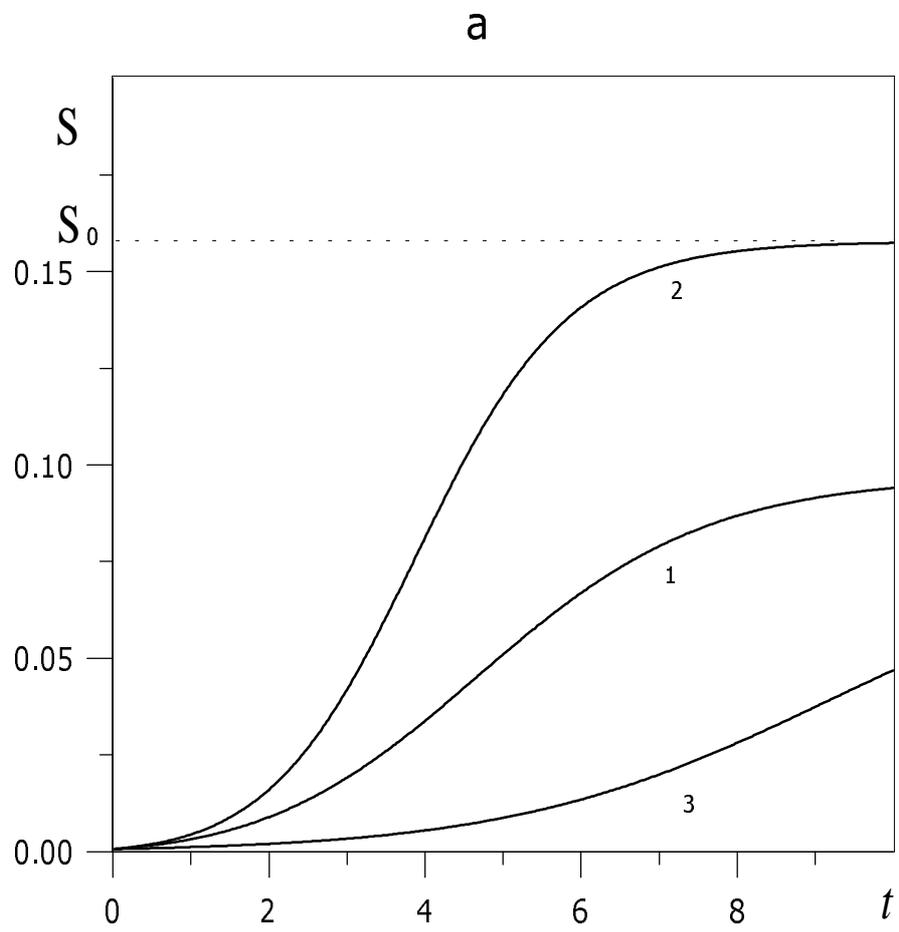
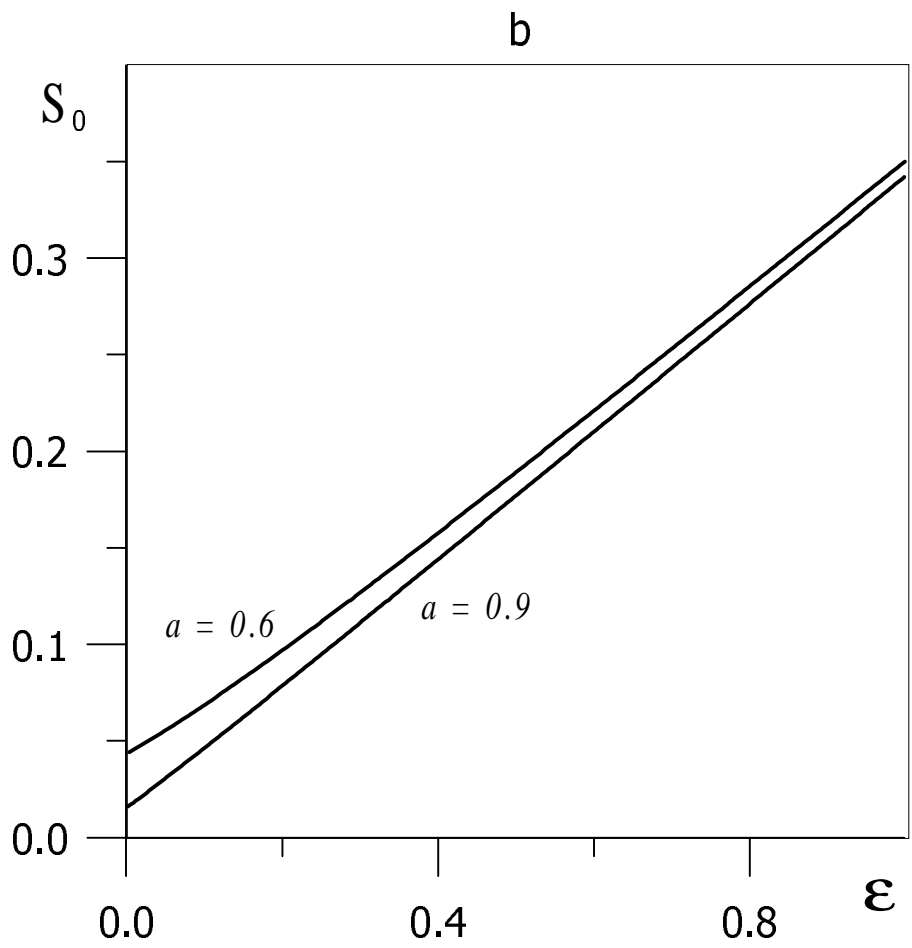
Fig.1. Structure factor behavior at $a>1/2$:\\ a) time dependencies
$S(t)$ (curves 1, 2, 3 correspond to $a=0.6$, $\varepsilon=0.2$;
$a=0.6$, $\varepsilon=0.4$; $a=0.9$, $\varepsilon=0.2$);\\
b) stationary point $S_0$ vs temperature $\varepsilon$ for several
values of exponent $a$.

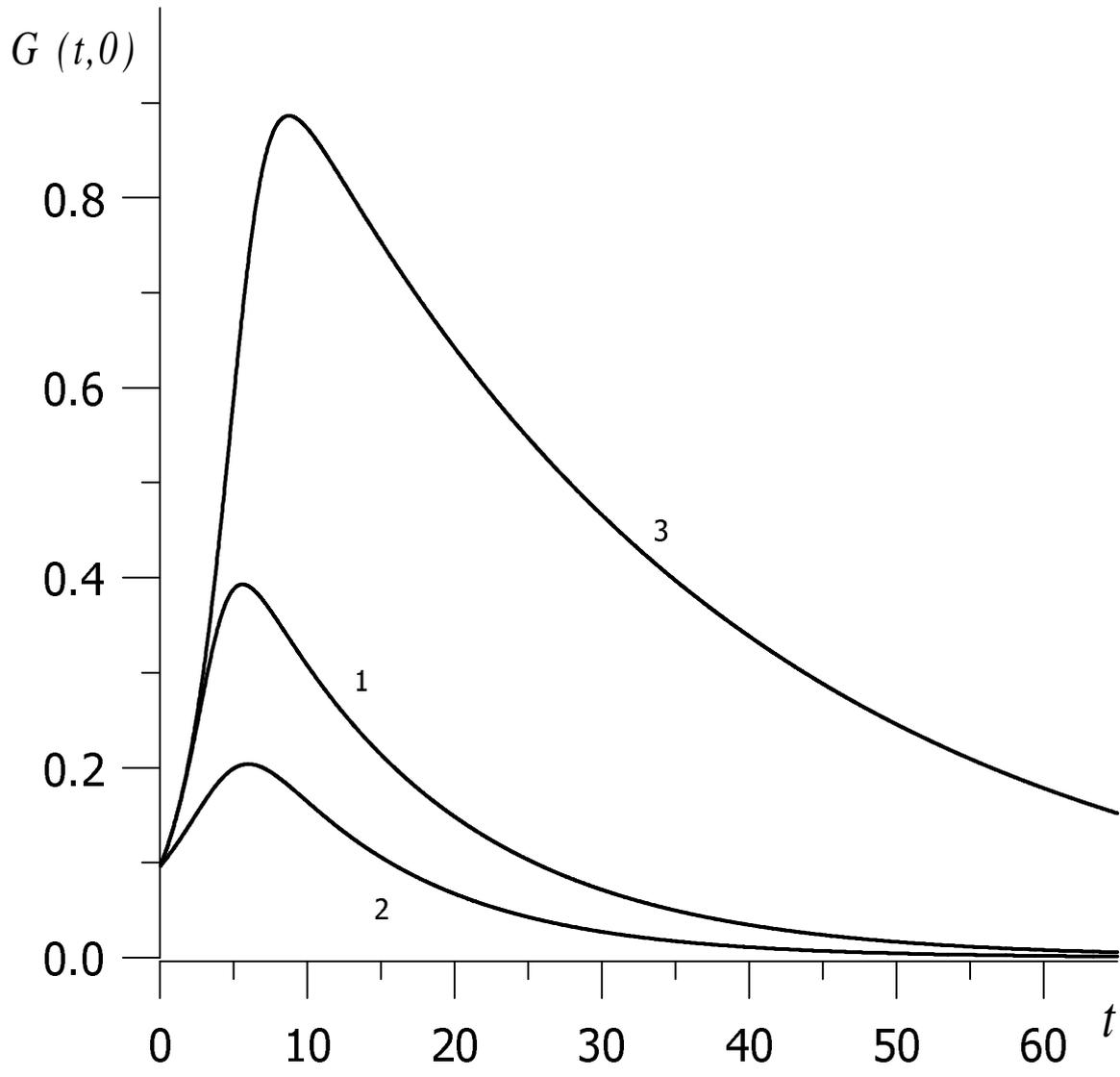
Fig.2. Correlator $G(t,0)$ vs time at $a>1/2$ for several exponent
$a$ and temperature $\varepsilon$ (curves 1, 2, 3 correspond to:
$a=0.6$, $\varepsilon=0.4$;
$a=0.6$, $\varepsilon=0.2$; $a=0.9$, $\varepsilon=0.4$).

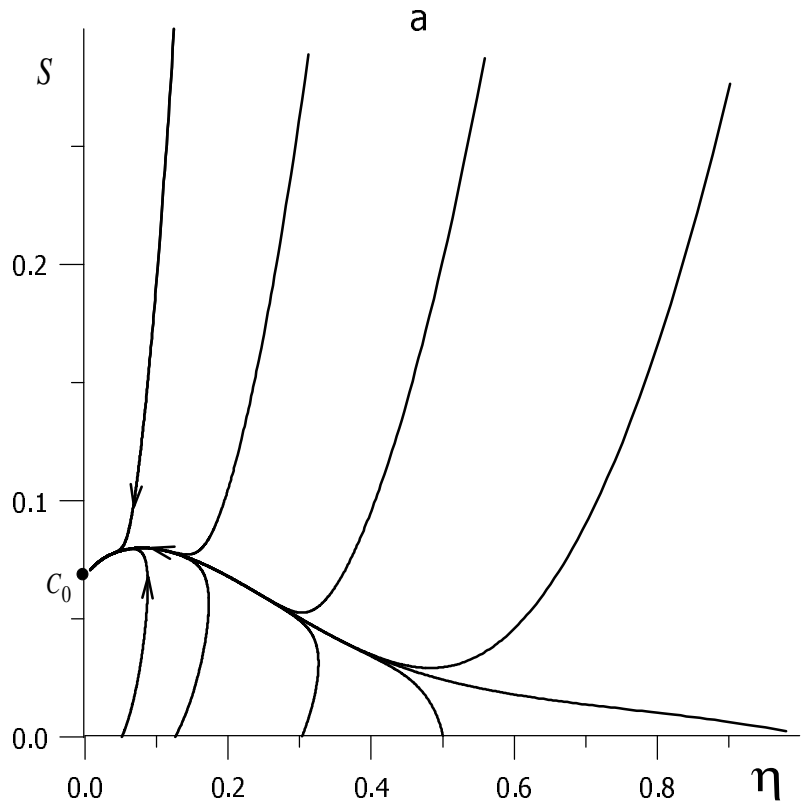
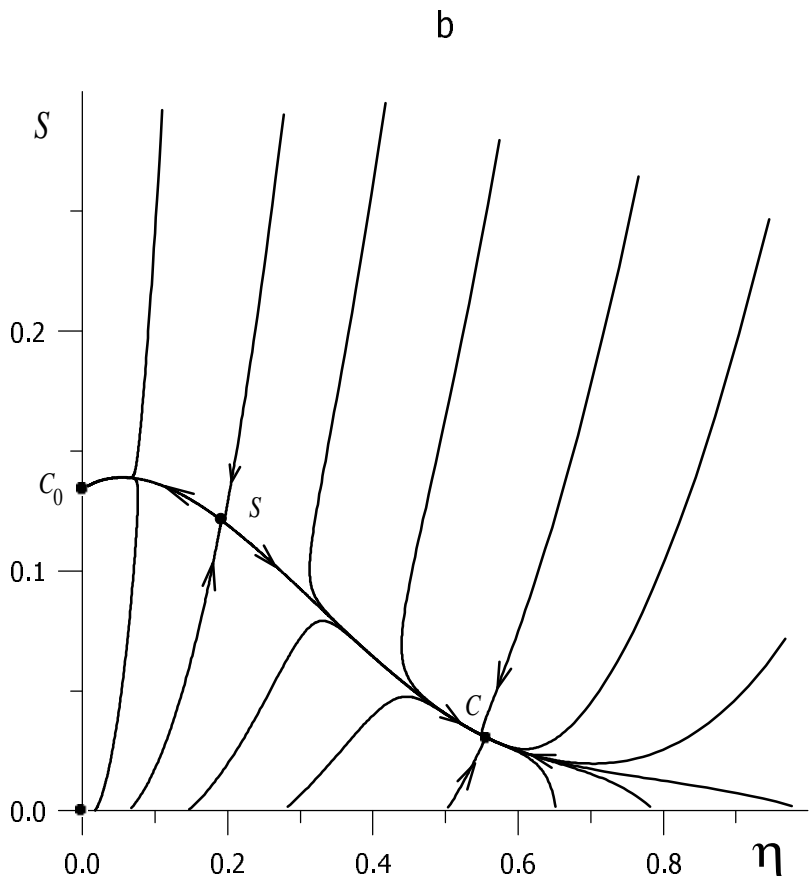
Fig.3. Phase portrait at $a>1/2$:\\
a) $a=0.3$, $\varepsilon=0.2$;\\
b) $a=0.3$, $\varepsilon=0.4$.

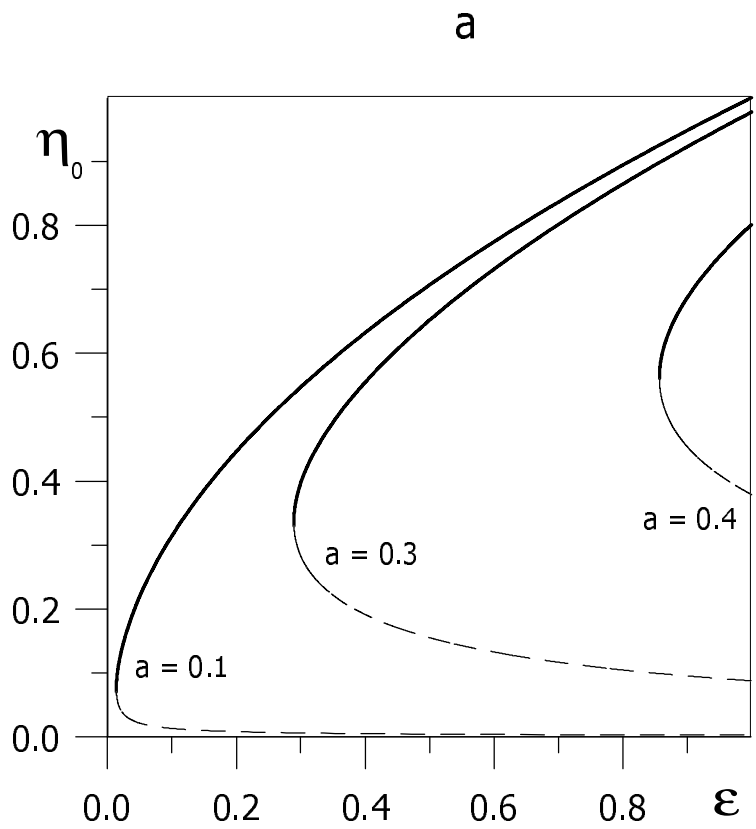
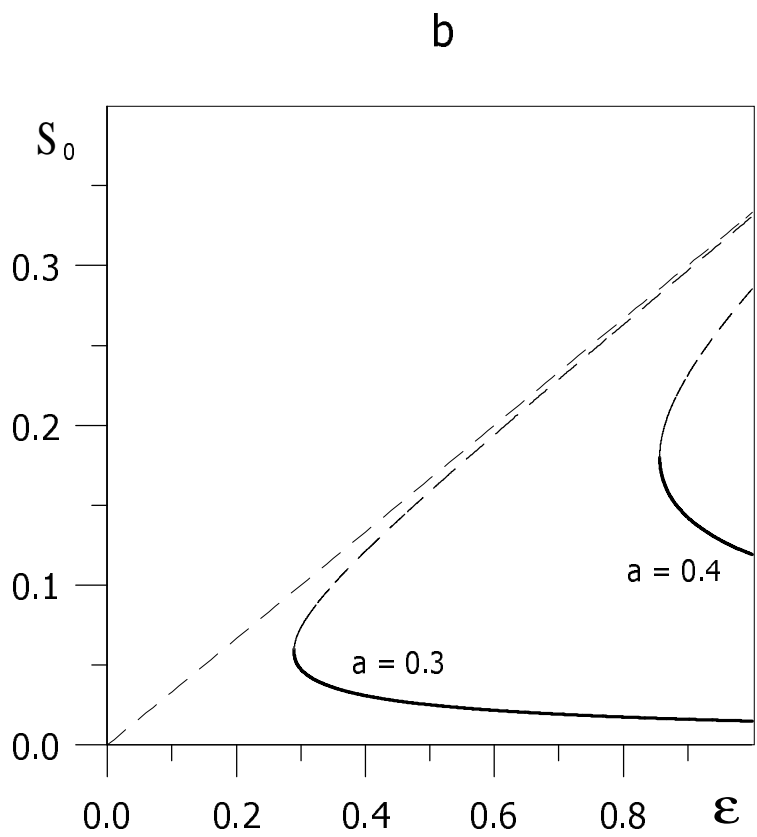
Fig.4. Stationary states of the system at $a>1/2$:\\
a)order parameter $\eta$ vs temperature $\varepsilon$ for several exponent
$a$;\\
b) structure factor $S$ vs temperature $\varepsilon$ for several exponent
$a$.

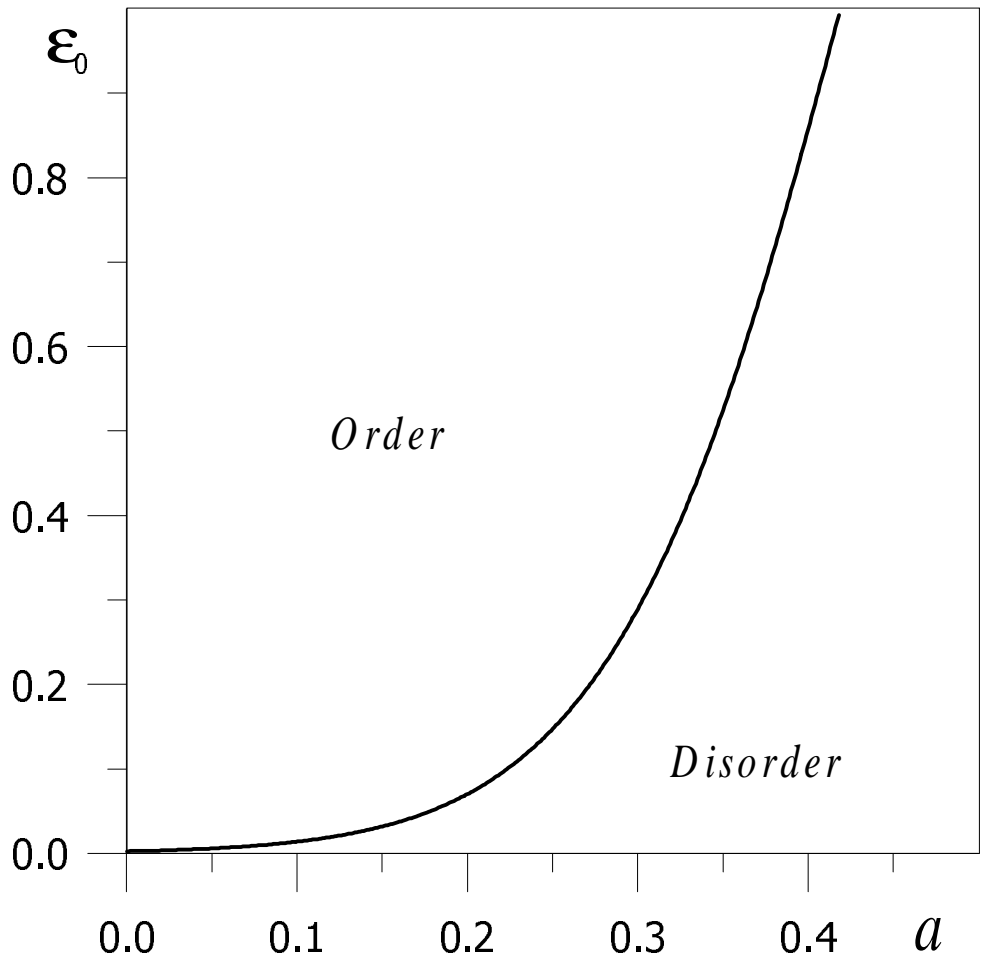
Fig.5. Phase diagram $\varepsilon_0(a)$.

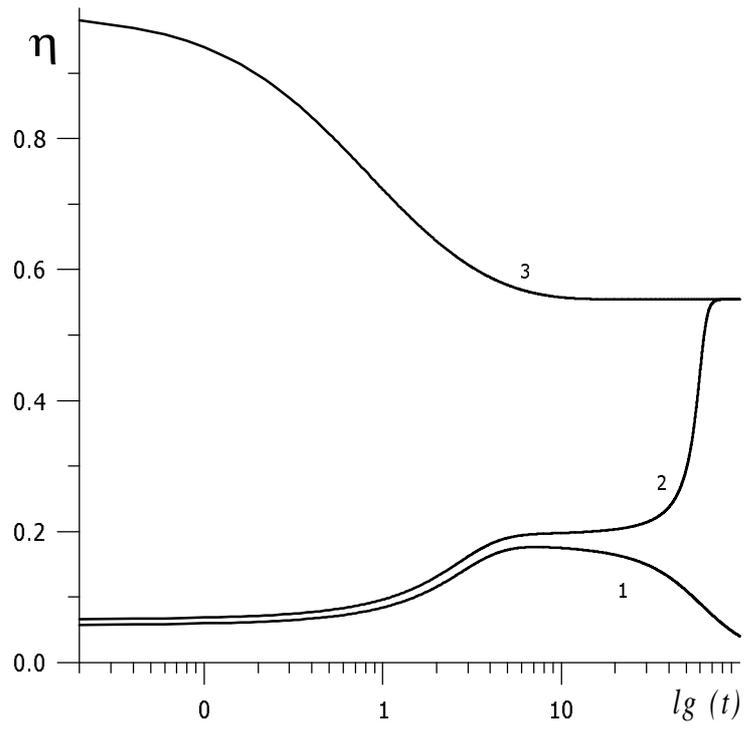
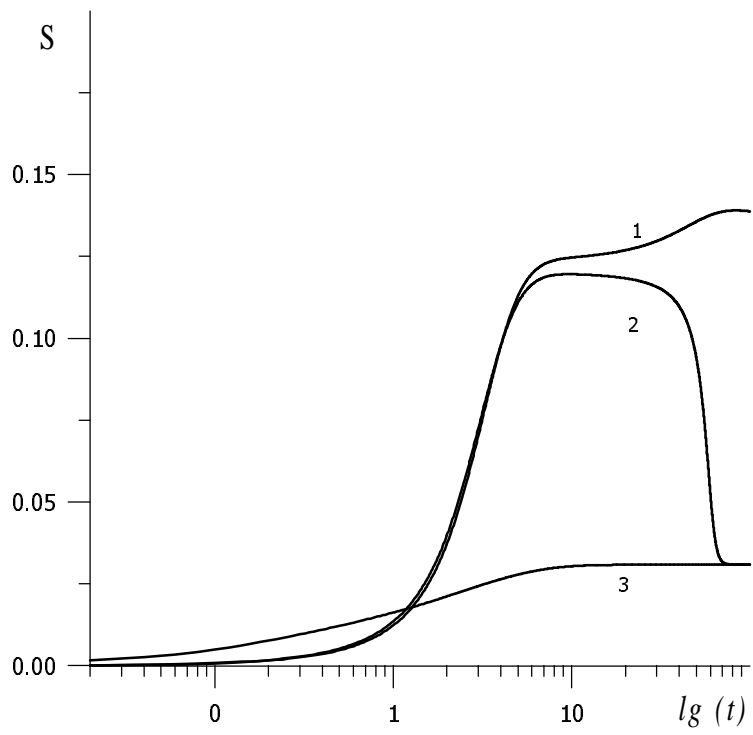
Fig.6. Time dependencies corresponded to different trajectories on
the phase portrait in Fig.3:\\
a) $\eta$ vs $\ln t$ at $a=0.3$, $\varepsilon=0.2$ and $S(0)=0$
(curves 1, 2, 3 correspond to $\eta(0)=0.057$, $\eta(0)=0.066$,
$\eta(0)=1.0$);\\
a) $S$ vs $\ln t$ at $a=0.3$, $\varepsilon=0.4$ and $S(0)=0$
(curves 1, 2, 3 correspond to $\eta(0)=0.057$, $\eta(0)=0.066$,
$\eta(0)=1.0$).

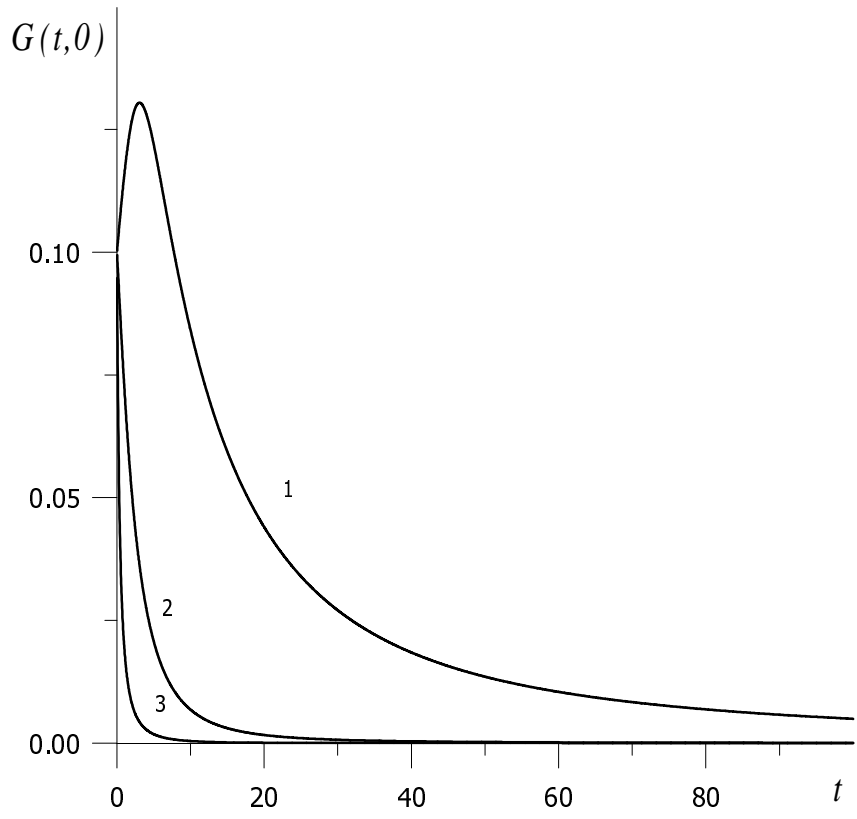
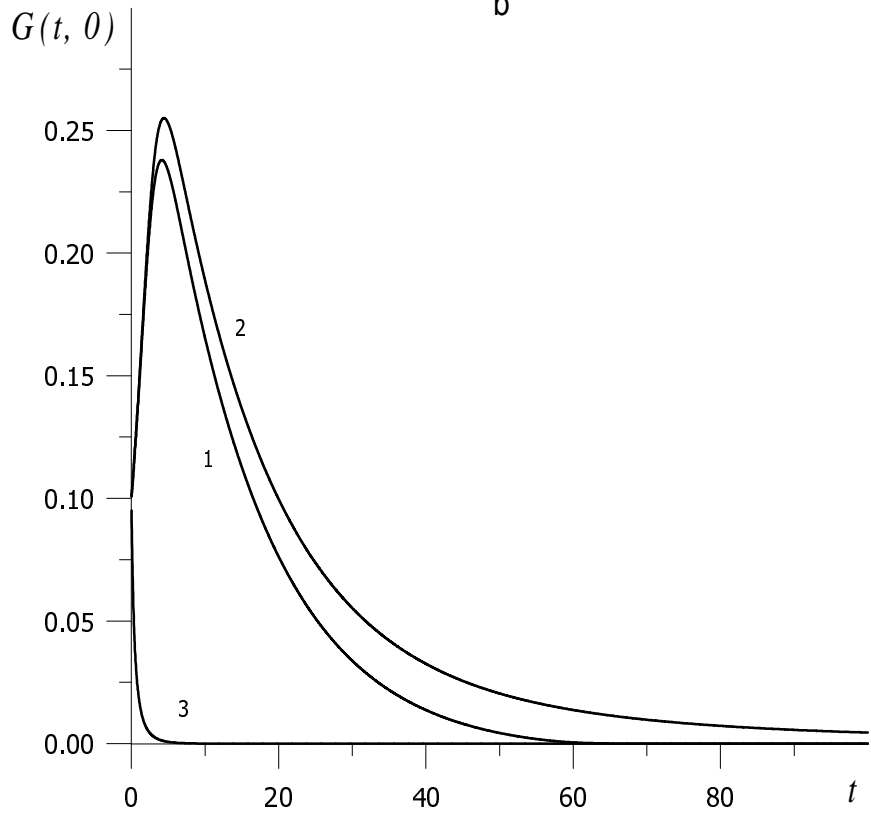
Fig.7. Correlator $G(t, 0)$ vs time $t$ at:\\
a) $a=0.3$, $\varepsilon=0.2$,  $S(0)=0$
(curves 1, 2, 3 correspond to $\eta(0)=0.098$, $\eta(0)=0.4$,
$\eta(0)=0.97$);\\
b) $a=0.3$, $\varepsilon=0.4$,  $S(0)=0$
(curves 1, 2, 3 correspond to $\eta(0)=0.057$, $\eta(0)=0.066$,
$\eta(0)=1.0$).

\end{document}